\documentclass[twocolumn,12pt]{iopart}

\usepackage{graphicx}

\begin{document}

\title[]{Complete hyperentangled Greenberger-Horne-Zeilinger state analysis for polarization and time-bin hyperentanglement}
\author{Zhi Zeng $^{1,2,3,^*}$}
\address{{$^1$Institute of Signal Processing and Transmission, Nanjing University of Posts and Telecommunications, Nanjing 210003, China}  \\
{$^2$Key Lab of Broadband Wireless Communication and Sensor Network Technology, Ministry of Education, Nanjing University of Posts and Telecommunications, Nanjing 210003, China}  \\
{$^3$School of Physics and Astronomy, Shanghai Jiao Tong University, Shanghai 200240, China}}
\eads{\mailto{zengzhiphy@yeah.net}}
\vspace{10pt}

\begin{abstract}
We present an efficient scheme for the complete analysis of hyperentangled Greenberger-Horne-Zeilinger (GHZ) state in polarization and time-bin degrees of freedom with two steps. First, the polarization GHZ state is distinguished completely and nondestructively, resorting to the controlled phase flip (CPF) gate constructed by the cavity-assisted interaction. Subsequently, the time-bin GHZ state is analyzed by using the preserved polarization entanglement. With the help of CPF gate and self-assisted mechanism, our scheme can be directly generalized to the complete $N$-photon hyperentangled GHZ state analysis, and it may have potential applications in the hyperentanglement-based quantum communication.
\end{abstract}



\section{Introduction}
Quantum entanglement is the unique phenomenon in the quantum world, and it has been widely used for many quantum communication tasks, such as quantum key distribution \cite{qkd}, quantum teleportation \cite{te}, quantum dense coding \cite{dense}, quantum secret sharing \cite{qss1,qss2} and quantum secure direct communication \cite{qsdc1,qsdc2,qsdc3}. Generally speaking, photon is considered to be the ideal candidate for the realization of quantum communication owing to its high-speed transmission and excellent low-noise properties. Several different degrees of freedom (DOFs) of photon can be utilized to encode quantum information, for example, the polarization, spatial-mode, frequency, time-bin and orbital angular momentum. Hyperentanglement, which is defined as the entanglement in more than one DOF, has attract much attention in recent years \cite{hyper1}. It can largely increase the channel capacity of quantum communication, and be useful for entangled state analysis \cite{BSA1,BSA2,GSA1,GSA2}, deterministic entanglement purification \cite{hyper2,hyper3,hyper4,hyper5} and hyper-parallel quantum computing \cite{QC1,QC2,QC3,QC4}. There are also some important works dealing with hyperentanglement itself, such as hyperentangled Bell state analysis (HBSA) \cite{HBSA1,HBSA2,HBSA3,HBSA4,HBSA5,HBSA6,HBSA7,HBSA8}, hyperentanglement purification and hyperentanglement concentration \cite{H1,H2,H3,H4,H5,H6}. 

The complete analysis of hyperentangled state is essential to high-capacity quantum information processing, and it cannot be accomplished with just linear optics \cite{no1,no2}. Although there are many proposals for the HBSA of two-photon system, only a few people give attention to the hyperentangled Greenberger-Horne-Zeilinger (GHZ) state analysis (HGSA) of multi-photon system. In 2012, Xia \emph{et al.} proposed the first complete HGSA scheme with the help of cross-Kerr nonlinearity \cite{HGSA1}. In 2013, Liu \emph{et al.} showed a deterministic scheme for the complete HGSA of three-photon system, resorting to the quantum dot spins in optical microcavities \cite{HGSA2}. In 2016, Li \emph{et al.} presented an efficient method for the $N$-photon HGSA, in which the discrimination process is simplified and the requirement on nonlinearity is reduced \cite{HGSA3}. In 2018, Zheng \emph{et al.} proposed the complete HGSA scheme with nitrogen-vacancy centers in microcavities \cite{HGSA4}. In 2020, the complete HGSA scheme with nonlinear interaction and auxiliary entanglement was also proposed \cite{HGSA5}.

The existing complete HGSA protocols for photon system in two DOFs are working with polarization and spatial-mode hyperentangled state, since this type of hyperentanglement can be prepared and manipulated with high fidelity at present. However, each photon requires more than one path during the transmission when the spatial-mode DOF is exploited, and it will lead to the requirement on extra quantum resource. Instead, the time-bin states of photon can be simply discriminated by the time of arrival. In this paper, we present a practical complete HGSA scheme for the polarization and time-bin hyperentanglement. In our scheme, the controlled phase flip (CPF) gate is used to determine the polarization GHZ state, and it can be realized with the cavity-assisted interaction. Then, the time-bin GHZ state is distinguished with the help of the preserved polarization entanglement. Our scheme is also suitable for the complete $N$-photon HGSA, and will be useful for the high-capacity quantum communication based on polarization-time-bin hyperentanglement. 

\section{The principle of CPF gate with cavity-assisted interaction}

\begin{figure}
\centering
\includegraphics*[width=0.7\textwidth]{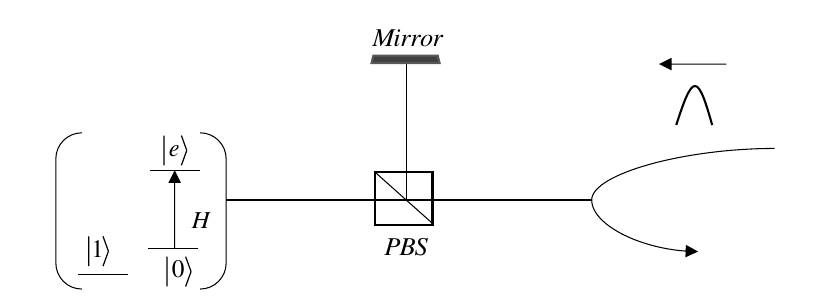}
\caption{Schematic diagram of the CPF gate with cavity-assisted interaction. An atom is trapped in the single-sided cavity, and it has a three-level structure. $|0\rangle$ and $|1\rangle$ are the two hyperfine levels in ground state, and $|e\rangle$ is the excited state. PBS is polarization beam splitter which transmits horizontal polarization photon and reflects the vertical one. Mirror is utilized to reflect the photon to its former path.}
\end{figure}

We briefly review the CPF gate between a photon and an atom inside single-sided optical microcavity \cite{CPF1}, as shown in Fig. 1. The confined atom has two ground states ($|0\rangle$ and $|1\rangle$) and one excited state ($|e\rangle$), and the transition $|0\rangle \rightarrow |e\rangle$ of the three-level atom is resonantly driven by $H$ polarization component of the input photonic state. When the condition $\kappa T \gg 1$ is satisfied, in which $\kappa$ is the decay rate of cavity and $T$ is the duration of input photon pulse, there are three possible outcomes after the interaction between photon and atom. When the polarization state of photon is $|H\rangle$, it will be reflected by the cavity with both its shape and phase unchanged if the atomic state is $|0\rangle$, and will be reflected with its shape almost unchanged but phase added by $\pi$ if the atomic state is $|1\rangle$. When the polarization state of photon is $|V\rangle$, it will be reflected by the mirror and no change will take place. Therefore, the CPF gate can be described by the following operator
\begin{eqnarray}
 U^{CPF} = e^{i\pi|1\rangle\langle1| \otimes |H\rangle\langle H|}.
\end{eqnarray}

\section{Complete HGSA for polarization and time-bin hyperentanglement}
The multi-photon hyperentangled GHZ state in polarization and time-bin DOFs can be written as
\begin{eqnarray}
|\Phi\rangle_{AB\cdots N} = |\Phi\rangle_{P} \otimes |\Phi\rangle_{T},
\end{eqnarray}
in which $A,B\cdots N$ represent the $N$ entangled photons. $|\Phi\rangle_{P}$ ($|\Phi\rangle_{T}$) is one of the $2^N$ polarization (time-bin) GHZ states,
\begin{eqnarray}
|\Phi^{\pm}_{ij\cdots k}\rangle_{P(T)} = \frac{1}{\sqrt 2} (|ij\cdots k\rangle \pm |\bar{i}\bar{j}\cdots \bar{k}\rangle)_{AB\cdots N}.
\end{eqnarray}
Here, $i,j\cdots k \in \{0,1\}$ and $m = 1-\bar{m} (m = i,j\cdots k)$. For the polarization DOF, $|0\rangle \equiv |H\rangle$ and $|1\rangle \equiv |V\rangle$. $|H\rangle$ and $|V\rangle$ denote the horizontal and vertical polarization states, respectively. For the time-bin DOF, $|0\rangle \equiv |S\rangle$ and $|1\rangle \equiv |L\rangle$. $|S\rangle$ and $|L\rangle$ denote two different time-bins, the early ($S$) and the later ($L$). Considering the two DOFs, there are $4^N$ hyperentangled GHZ states, which will be distinguished in this section. We first describe our scheme for the complete three-photon HGSA in detail, and then generalize it to the complete $N$-photon HGSA directly.

When $N = 3$, one of the 64 mutually orthogonal hyperentangled GHZ states can be written as
\begin{eqnarray}
|\Phi^{+}_{000}\rangle_{P} \otimes |\Phi^{+}_{000}\rangle_{T} = \frac{1}{\sqrt 2} (|HHH\rangle + |VVV\rangle) \otimes \frac{1}{\sqrt 2} (|SSS\rangle + |LLL\rangle)_{ABC}.  \nonumber\\ 
\end{eqnarray}

\begin{figure}
\centering
\includegraphics*[width=0.8\textwidth]{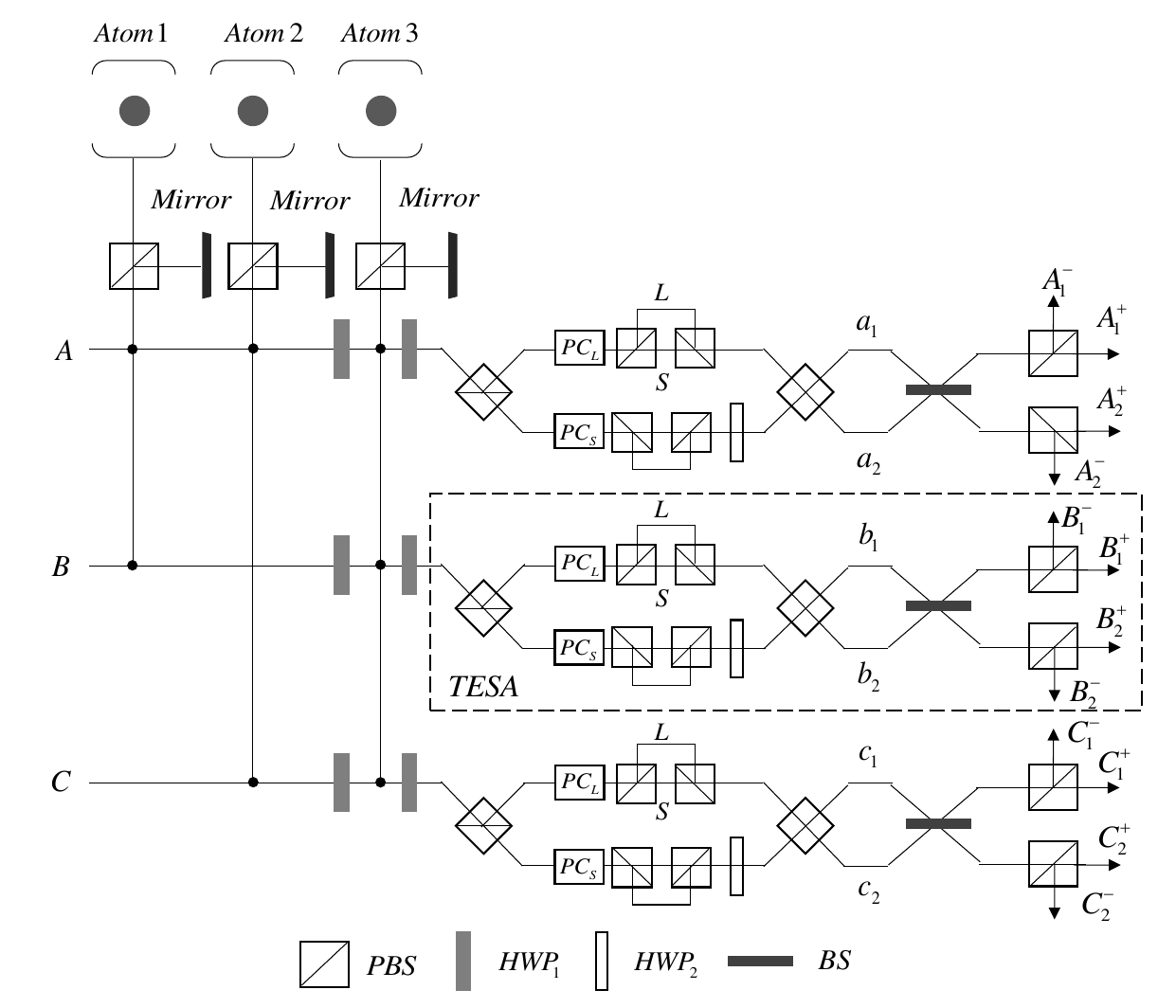}
\caption{Schematic diagram of our complete HGSA scheme for three-photon system. HWP$_1$ is a half wave plate, which can perform the Hadamard operation [$|H\rangle \rightarrow \frac{1}{\sqrt 2} (|H\rangle + |V\rangle)$ and $|V\rangle \rightarrow \frac{1}{\sqrt 2} (|H\rangle - |V\rangle)$] on the polarization state of photon. PC$_S$ (PC$_L$) is the Pockels cell which can effect the bit-flip operation on polarization state when the $S$ ($L$) component is present. The unbalanced interferometer composed of two PBSs can be used to adjust the time-bin state of photon. HWP$_2$ is another half wave plate, which can perform the bit-flip operation on polarization state. BS is the 50:50 beam splitter and it can perform the Hadamard operation [$|x_1\rangle \rightarrow \frac{1}{\sqrt 2} (|x_1\rangle + |x_2\rangle)$ and $|x_2\rangle \rightarrow \frac{1}{\sqrt 2} (|x_1\rangle - |x_2\rangle) (x = a, b, c)$] on the path state of photon. The dashed rectangle represents the process of time-bin entangled state analysis (TESA), which can be achieved by using the preserved polarization entanglement. With the help of the three atoms in cavities and single photon detectors, the 64 hyperentangled GHZ states in polarization and time-bin DOFs can be completely distinguished.}
\end{figure}

The setup of our complete three-photon HGSA scheme for polarization and time-bin hyperentanglement is shown in Fig. 2. The quantum states of atoms are defined as $|+\rangle = (|0\rangle + |1\rangle)/\sqrt{2}$ and $|-\rangle = (|0\rangle - |1\rangle)/\sqrt{2}$, and all of the three atoms are initially prepared in state $|+\rangle$. After photons $A$, $B$ and $C$ interact with the atom 1 and atom 2 in single-sided cavity, the state of the collective system composed of hyperentanglement and two atoms evolves as
\begin{eqnarray}
|\Phi^{\pm}_{000}\rangle_P|\Phi\rangle_T \otimes |+\rangle_1|+\rangle_2 \rightarrow |\Phi^{\pm}_{000}\rangle_P|\Phi\rangle_T \otimes |+\rangle_1|+\rangle_2, \nonumber\\
|\Phi^{\pm}_{001}\rangle_P|\Phi\rangle_T \otimes |+\rangle_1|+\rangle_2 \rightarrow |\Phi^{\pm}_{001}\rangle_P|\Phi\rangle_T \otimes |+\rangle_1|-\rangle_2, \nonumber\\
|\Phi^{\pm}_{010}\rangle_P|\Phi\rangle_T \otimes |+\rangle_1|+\rangle_2 \rightarrow |\Phi^{\pm}_{010}\rangle_P|\Phi\rangle_T \otimes |-\rangle_1|+\rangle_2, \nonumber\\
|\Phi^{\pm}_{100}\rangle_P|\Phi\rangle_T \otimes |+\rangle_1|+\rangle_2 \rightarrow |\Phi^{\pm}_{100}\rangle_P|\Phi\rangle_T \otimes |-\rangle_1|-\rangle_2.
\end{eqnarray}
We can find that the parity information of polarization entanglement can be obtained with the help of the measurement on two atoms. After the three photons pass through HWPs, interact with the atom 3 and pass through the HWPs again, the evolution of the collective system can be expressed as
\begin{eqnarray}
|\Phi^{+}_{ijk}\rangle_P|\Phi\rangle_T \otimes |+\rangle_3 \rightarrow |\Phi^{+}_{ijk}\rangle_P|\Phi\rangle_T \otimes |-\rangle_3, \nonumber\\
|\Phi^{-}_{ijk}\rangle_P|\Phi\rangle_T \otimes |+\rangle_3 \rightarrow |\Phi^{-}_{ijk}\rangle_P|\Phi\rangle_T \otimes |+\rangle_3.
\end{eqnarray}
$|\Phi^{+}_{ijk}\rangle_P$ represents the polarization GHZ state with "+" phase information ($|\Phi^{+}_{000}\rangle_P$, $|\Phi^{+}_{001}\rangle_P$, $|\Phi^{+}_{010}\rangle_P$ or $|\Phi^{+}_{100}\rangle_P$), while $|\Phi^{-}_{ijk}\rangle_P$ represents the polarization GHZ state with "-" phase information ($|\Phi^{-}_{000}\rangle_P$, $|\Phi^{-}_{001}\rangle_P$, $|\Phi^{-}_{010}\rangle_P$ or $|\Phi^{-}_{100}\rangle_P$). Thus, the discrimination of $|\Phi^{+}_{ijk}\rangle_P$ and $|\Phi^{-}_{ijk}\rangle_P$ can be realized by the detection of atom 3. The relationship between the initial states and their corresponding final atomic states is summarized in Table 1.

\begin{table}
\centering\caption{Corresponding relations between the initial states and the final atomic states.}
\begin{tabular}{cc ccccccccccc}
\hline
Initial state & & & & Atom 1 & & & & Atom 2 & & & & Atom 3 \\
\hline
$|\Phi^{+}_{000}\rangle_P|\Phi\rangle_T$ &&&& $|+\rangle_1$ &&&& $|+\rangle_2$ &&&& $|-\rangle_3$ \\
$|\Phi^{-}_{000}\rangle_P|\Phi\rangle_T$ &&&& $|+\rangle_1$ &&&& $|+\rangle_2$ &&&& $|+\rangle_3$ \\
$|\Phi^{+}_{001}\rangle_P|\Phi\rangle_T$ &&&& $|+\rangle_1$ &&&& $|-\rangle_2$ &&&& $|-\rangle_3$ \\
$|\Phi^{-}_{001}\rangle_P|\Phi\rangle_T$ &&&& $|+\rangle_1$ &&&& $|-\rangle_2$ &&&& $|+\rangle_3$ \\
$|\Phi^{+}_{010}\rangle_P|\Phi\rangle_T$ &&&& $|-\rangle_1$ &&&& $|+\rangle_2$ &&&& $|-\rangle_3$ \\
$|\Phi^{-}_{010}\rangle_P|\Phi\rangle_T$ &&&& $|-\rangle_1$ &&&& $|+\rangle_2$ &&&& $|+\rangle_3$ \\
$|\Phi^{+}_{100}\rangle_P|\Phi\rangle_T$ &&&& $|-\rangle_1$ &&&& $|-\rangle_2$ &&&& $|-\rangle_3$ \\
$|\Phi^{-}_{100}\rangle_P|\Phi\rangle_T$ &&&& $|-\rangle_1$ &&&& $|-\rangle_2$ &&&& $|+\rangle_3$ \\
\hline
\end{tabular}
\end{table}

The polarization GHZ states are already separated with the help of three atoms in cavities, completely and nondestructively. After that, the preserved polarization entanglement can be used as an ancillary to distinguish the time-bin GHZ states, which is known as the hyperentanglement-assisted principle. For the description, assume that the preserved polarization GHZ state is $|\Phi^{+}_{001}\rangle_P$. It should be noted that the other seven polarization GHZ states are also useful for the discrimination of time-bin entanglement. After the three photons pass through the TESA in Fig. 2, the hyperentanglement in polarization and time-bin DOFs will evolve as
\begin{eqnarray}
|\Phi^{+}_{001}\rangle_{P}|\Phi^{+}_{000}\rangle_{T} \rightarrow &&\frac{1}{2{\sqrt 2}} (|HHV\rangle + |VVH\rangle) \nonumber \\ &&\otimes (|a_{1}b_{1}c_{1}\rangle + |a_{1}b_{2}c_{2}\rangle + |a_{2}b_{1}c_{2}\rangle +|a_{2}b_{2}c_{1}\rangle)_{ABC}, \nonumber \\
|\Phi^{+}_{001}\rangle_{P}|\Phi^{-}_{000}\rangle_{T} \rightarrow &&\frac{1}{2{\sqrt 2}} (|HHV\rangle + |VVH\rangle) \nonumber \\ &&\otimes (|a_{1}b_{1}c_{2}\rangle + |a_{1}b_{2}c_{1}\rangle + |a_{2}b_{1}c_{1}\rangle +|a_{2}b_{2}c_{2}\rangle)_{ABC},  \nonumber \\
|\Phi^{+}_{001}\rangle_{P}|\Phi^{+}_{001}\rangle_{T} \rightarrow &&\frac{1}{2{\sqrt 2}} (|HHH\rangle + |VVV\rangle) \nonumber \\ &&\otimes (|a_{1}b_{1}c_{1}\rangle - |a_{1}b_{2}c_{2}\rangle - |a_{2}b_{1}c_{2}\rangle +|a_{2}b_{2}c_{1}\rangle)_{ABC}, \nonumber \\
|\Phi^{+}_{001}\rangle_{P}|\Phi^{-}_{001}\rangle_{T} \rightarrow &&\frac{1}{2{\sqrt 2}} (|HHH\rangle + |VVV\rangle) \nonumber \\ &&\otimes (|a_{1}b_{1}c_{2}\rangle - |a_{1}b_{2}c_{1}\rangle - |a_{2}b_{1}c_{1}\rangle +|a_{2}b_{2}c_{2}\rangle)_{ABC},  \nonumber \\
|\Phi^{+}_{001}\rangle_{P}|\Phi^{+}_{010}\rangle_{T} \rightarrow &&\frac{1}{2{\sqrt 2}} (|HVV\rangle + |VHH\rangle) \nonumber \\ &&\otimes (|a_{1}b_{1}c_{1}\rangle - |a_{1}b_{2}c_{2}\rangle + |a_{2}b_{1}c_{2}\rangle -|a_{2}b_{2}c_{1}\rangle)_{ABC}, \nonumber \\
|\Phi^{+}_{001}\rangle_{P}|\Phi^{-}_{010}\rangle_{T} \rightarrow &&\frac{1}{2{\sqrt 2}} (|HVV\rangle + |VHH\rangle) \nonumber \\ &&\otimes (|a_{1}b_{1}c_{2}\rangle - |a_{1}b_{2}c_{1}\rangle + |a_{2}b_{1}c_{1}\rangle -|a_{2}b_{2}c_{2}\rangle)_{ABC},  \nonumber \\
|\Phi^{+}_{001}\rangle_{P}|\Phi^{+}_{100}\rangle_{T} \rightarrow &&\frac{1}{2{\sqrt 2}} (|VHV\rangle + |HVH\rangle) \nonumber \\ &&\otimes (|a_{1}b_{1}c_{1}\rangle + |a_{1}b_{2}c_{2}\rangle - |a_{2}b_{1}c_{2}\rangle -|a_{2}b_{2}c_{1}\rangle)_{ABC}, \nonumber \\
|\Phi^{+}_{001}\rangle_{P}|\Phi^{-}_{100}\rangle_{T} \rightarrow &&\frac{1}{2{\sqrt 2}} (|VHV\rangle + |HVH\rangle) \nonumber \\ &&\otimes (|a_{1}b_{1}c_{2}\rangle + |a_{1}b_{2}c_{1}\rangle - |a_{2}b_{1}c_{1}\rangle -|a_{2}b_{2}c_{2}\rangle)_{ABC}.
\end{eqnarray}
The above eight states can be distinguished by the measurement results of single photon detectors, with which the initial 64 hyperentangled GHZ states in polarization and time-bin DOFs can be classified into eight groups, as shown in Table 2. In each group, there are eight states, which can be determinately identified through the Table 1. Thus, the complete three-photon HGSA can be accomplished by using Table 1 and Table 2 simultaneously.

\begin{table}
\centering\caption{The initial 64 hyperentangled GHZ states in two DOFs can be classified into eight groups based on the measurement results of single photon detectors.}
\begin{tabular}{cc ccccccccccc}
\hline
Group &  Initial states \\
\hline
$1$ & $|\Phi^{+}_{000}\rangle_P|\Phi^{+}_{000}\rangle_T,|\Phi^{+}_{001}\rangle_P|\Phi^{+}_{001}\rangle_T,$
      $|\Phi^{+}_{010}\rangle_P|\Phi^{+}_{010}\rangle_T,|\Phi^{+}_{100}\rangle_P|\Phi^{+}_{100}\rangle_T,$ \\ &
      $|\Phi^{-}_{000}\rangle_P|\Phi^{-}_{000}\rangle_T,|\Phi^{-}_{001}\rangle_P|\Phi^{-}_{001}\rangle_T,$
      $|\Phi^{-}_{010}\rangle_P|\Phi^{-}_{010}\rangle_T,|\Phi^{-}_{100}\rangle_P|\Phi^{-}_{100}\rangle_T.$ \\

$2$ & $|\Phi^{+}_{000}\rangle_P|\Phi^{-}_{000}\rangle_T,|\Phi^{+}_{001}\rangle_P|\Phi^{-}_{001}\rangle_T,$
      $|\Phi^{+}_{010}\rangle_P|\Phi^{-}_{010}\rangle_T,|\Phi^{+}_{100}\rangle_P|\Phi^{-}_{100}\rangle_T,$ \\ &
      $|\Phi^{-}_{000}\rangle_P|\Phi^{+}_{000}\rangle_T,|\Phi^{-}_{001}\rangle_P|\Phi^{+}_{001}\rangle_T,$
      $|\Phi^{-}_{010}\rangle_P|\Phi^{+}_{010}\rangle_T,|\Phi^{-}_{100}\rangle_P|\Phi^{+}_{100}\rangle_T.$ \\

$3$ & $|\Phi^{+}_{000}\rangle_P|\Phi^{+}_{001}\rangle_T,|\Phi^{+}_{001}\rangle_P|\Phi^{+}_{000}\rangle_T,$
      $|\Phi^{+}_{010}\rangle_P|\Phi^{+}_{100}\rangle_T,|\Phi^{+}_{100}\rangle_P|\Phi^{+}_{010}\rangle_T,$ \\ &
      $|\Phi^{-}_{000}\rangle_P|\Phi^{-}_{001}\rangle_T,|\Phi^{-}_{001}\rangle_P|\Phi^{-}_{000}\rangle_T,$
      $|\Phi^{-}_{010}\rangle_P|\Phi^{-}_{100}\rangle_T,|\Phi^{-}_{100}\rangle_P|\Phi^{-}_{010}\rangle_T.$ \\

$4$ & $|\Phi^{+}_{000}\rangle_P|\Phi^{-}_{001}\rangle_T,|\Phi^{+}_{001}\rangle_P|\Phi^{-}_{000}\rangle_T,$
      $|\Phi^{+}_{010}\rangle_P|\Phi^{-}_{100}\rangle_T,|\Phi^{+}_{100}\rangle_P|\Phi^{-}_{010}\rangle_T,$ \\ &
      $|\Phi^{-}_{000}\rangle_P|\Phi^{+}_{001}\rangle_T,|\Phi^{-}_{001}\rangle_P|\Phi^{+}_{000}\rangle_T,$
      $|\Phi^{-}_{010}\rangle_P|\Phi^{+}_{100}\rangle_T,|\Phi^{-}_{100}\rangle_P|\Phi^{+}_{010}\rangle_T.$ \\

$5$ & $|\Phi^{+}_{000}\rangle_P|\Phi^{+}_{010}\rangle_T,|\Phi^{+}_{001}\rangle_P|\Phi^{+}_{100}\rangle_T,$
      $|\Phi^{+}_{010}\rangle_P|\Phi^{+}_{000}\rangle_T,|\Phi^{+}_{100}\rangle_P|\Phi^{+}_{001}\rangle_T,$ \\ &
      $|\Phi^{-}_{000}\rangle_P|\Phi^{-}_{010}\rangle_T,|\Phi^{-}_{001}\rangle_P|\Phi^{-}_{100}\rangle_T,$
      $|\Phi^{-}_{010}\rangle_P|\Phi^{-}_{000}\rangle_T,|\Phi^{-}_{100}\rangle_P|\Phi^{-}_{001}\rangle_T.$ \\

$6$ & $|\Phi^{+}_{000}\rangle_P|\Phi^{-}_{010}\rangle_T,|\Phi^{+}_{001}\rangle_P|\Phi^{-}_{100}\rangle_T,$
      $|\Phi^{+}_{010}\rangle_P|\Phi^{-}_{000}\rangle_T,|\Phi^{+}_{100}\rangle_P|\Phi^{-}_{001}\rangle_T,$ \\ &
      $|\Phi^{-}_{000}\rangle_P|\Phi^{+}_{010}\rangle_T,|\Phi^{-}_{001}\rangle_P|\Phi^{+}_{100}\rangle_T,$
      $|\Phi^{-}_{010}\rangle_P|\Phi^{+}_{000}\rangle_T,|\Phi^{-}_{100}\rangle_P|\Phi^{+}_{001}\rangle_T.$ \\

$7$ & $|\Phi^{+}_{000}\rangle_P|\Phi^{+}_{100}\rangle_T,|\Phi^{+}_{001}\rangle_P|\Phi^{+}_{010}\rangle_T,$
      $|\Phi^{+}_{010}\rangle_P|\Phi^{+}_{001}\rangle_T,|\Phi^{+}_{100}\rangle_P|\Phi^{+}_{000}\rangle_T,$ \\ &
      $|\Phi^{-}_{000}\rangle_P|\Phi^{-}_{100}\rangle_T,|\Phi^{-}_{001}\rangle_P|\Phi^{-}_{010}\rangle_T,$
      $|\Phi^{-}_{010}\rangle_P|\Phi^{-}_{001}\rangle_T,|\Phi^{-}_{100}\rangle_P|\Phi^{-}_{000}\rangle_T.$ \\

$8$ & $|\Phi^{+}_{000}\rangle_P|\Phi^{-}_{100}\rangle_T,|\Phi^{+}_{001}\rangle_P|\Phi^{-}_{010}\rangle_T,$
      $|\Phi^{+}_{010}\rangle_P|\Phi^{-}_{001}\rangle_T,|\Phi^{+}_{100}\rangle_P|\Phi^{-}_{000}\rangle_T,$ \\ &
      $|\Phi^{-}_{000}\rangle_P|\Phi^{+}_{100}\rangle_T,|\Phi^{-}_{001}\rangle_P|\Phi^{+}_{010}\rangle_T,$
      $|\Phi^{-}_{010}\rangle_P|\Phi^{+}_{001}\rangle_T,|\Phi^{-}_{100}\rangle_P|\Phi^{+}_{000}\rangle_T.$ \\
\hline
\end{tabular}
\end{table}

\begin{figure}
\centering
\includegraphics*[width=0.7\textwidth]{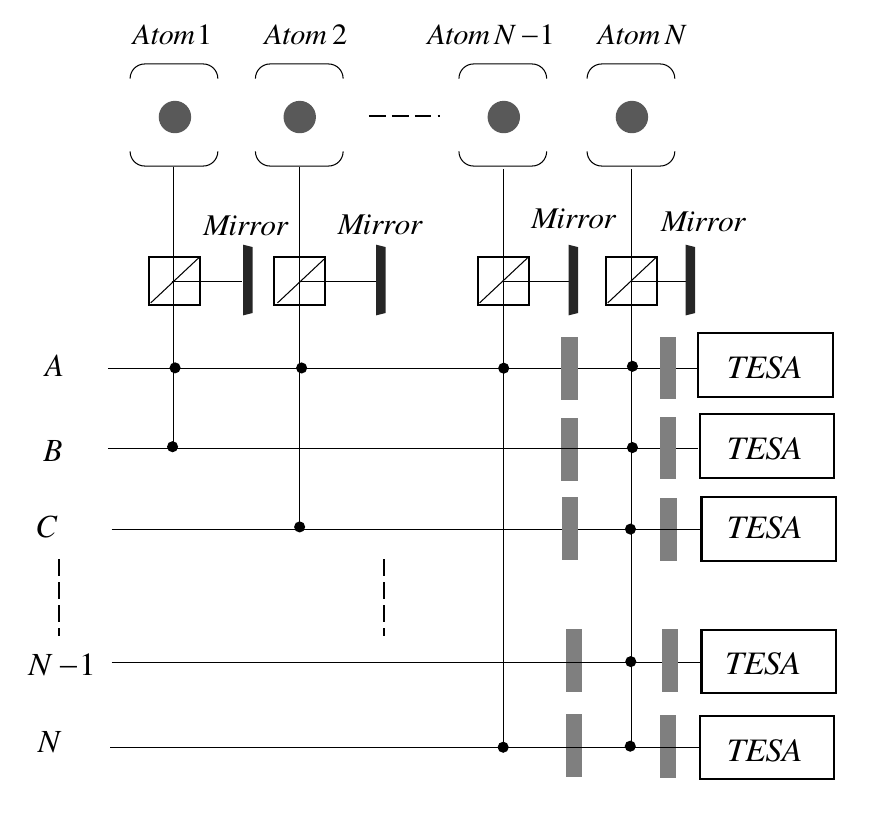}
\caption{Schematic diagram of our complete HGSA scheme for $N$-photon system. With the help of the $N$ atoms in cavities and single photon detectors, the $4^N$ hyperentangled GHZ states in polarization and time-bin DOFs can be completely distinguished.}
\end{figure}

Our scheme can be extended to the complete $N$-photon HGSA assisted by more atoms in cavities, as shown in Fig. 3. The $N$ atoms initially prepared in state $|+\rangle$ are utilized to distinguish the polarization GHZ state in hyperentanglement, the process of which can be expressed as
\begin{eqnarray}
&&|\Phi^{+}_{ijp\cdots qk}\rangle_{P}|\Phi\rangle_{T}|+\rangle_1|+\rangle_2\cdots|+\rangle_{N-2}             |+\rangle_{N-1} |+\rangle_{N} \nonumber \\
&&\rightarrow |\Phi^{+}_{ijp\cdots qk}\rangle_{P}|\Phi\rangle_{T}|\widetilde{i\oplus j}\rangle_1|\widetilde{i\oplus p}\rangle_2\cdots|\widetilde{i\oplus q}\rangle_{N-2}|\widetilde{i\oplus k}\rangle_{N-1}  |-\rangle_{N},
\end{eqnarray}
and
\begin{eqnarray}
&&|\Phi^{-}_{ijp\cdots qk}\rangle_{P}|\Phi\rangle_{T}|+\rangle_1|+\rangle_2\cdots|+\rangle_{N-2}             |+\rangle_{N-1} |+\rangle_{N} \nonumber \\
&&\rightarrow |\Phi^{-}_{ijp\cdots qk}\rangle_{P}|\Phi\rangle_{T}|\widetilde{i\oplus j}\rangle_1|\widetilde{i\oplus p}\rangle_2\cdots|\widetilde{i\oplus q}\rangle_{N-2}|\widetilde{i\oplus k}\rangle_{N-1}   |+\rangle_{N}.
\end{eqnarray}
Here, we define the atomic states as $|\widetilde0\rangle=|+\rangle$ and $|\widetilde1\rangle=|-\rangle$. After the measurement on the $N$ atoms, the polarization GHZ state can be nondestructively determined. Then, the preserved polarization entanglement will be used as the ancillary to distinguish the time-bin GHZ state, and this process is similar to the Equation (7). With these two independent steps, the complete $N$-photon HGSA for polarization-time-bin hyperentanglement is accomplished.

\section{Discussion and summary}
In the past years, the schemes for quantum state analysis have been accomplished experimentally, which can benefit the practical quantum communication a lot \cite{AD1,AD2,AD3,AD4}. In 2006, Schuck \emph{et al.} deterministically distinguished the four polarization Bell states of two photons, using the hyperentanglement in polarization and time-bin DOFs \cite{AD1}. In 2008, Barreiro \emph{et al.} realized the polarization Bell state analysis assisted by the orbital angular momentum DOF, which can be helpful for the dense coding with spin–orbit encoded photons \cite{AD3}. In 2017, Williams \emph{et al.} reported the first demonstration of superdense coding over optical fiber links, resorting to the complete Bell state analysis enabled by time-polarization hyperentanglement \cite{AD4}. Besides Bell state analysis, the implementation of GHZ state analysis for multi-photon system is also important to quantum communication \cite{AD5,AD6,AD7,AD8}. For example, in 2009, Lu \emph{et al.} reported the first experimental demonstration of GHZ entanglement swapping, in which the GHZ state analysis was required \cite{AD5}. In 2015, Fu \emph{et al.} experimental demonstration the measurement-device-independent multiparty quantum communication proposal, in which the GHZ state analysis has been utilized \cite{AD6}. Also, the generation of hyperentangled GHZ state for multi-photon system in two or three DOFs have been realized in experiment \cite{AD9,AD10}. 

In the presented scheme, the atom-cavity system is exploited to construct the CPF gate, which plays an important role in our complete HGSA scheme. Therefore, it is necessary for us to discuss the feasibility of CPF gate with the current technology. In 2004, Duan \emph{et al.} proposed the CPF gate in theory, and their numerical simulations showed that the CPF gate is robust to the practical noise and experimental imperfections in the cavity-QED setups \cite{CPF1}. Considering a neutral atom trapped in the Fabry-Perot cavity with $(g, \kappa, \Gamma)/2\pi \approx (25, 8, 5.2)MHz$, one may calculate the gate fidelity $F$ to be 0.999 if $T = 240/\kappa$ and the error probability $P_s = 0.032$ due to the spontaneous emission \cite{CPF1}. In the same year, Xiao \emph{at al.} proposed a scheme to realize the CPF gate between two rare-earth ions embedded in cavity, and their numerical simulations showed that the CPF gate is robust and scalable with high fidelity and low error rate \cite{CPF2}. In the case of a rare-earth ion embedded in a silica-microsphere cavity with $(g, \kappa, \Gamma)/2\pi \approx (10^{3}, 32, 10^{-3})MHz$ and $T = 3\mu s$, $F$ can reach 99.998\% and $P_s$ is about $10^{-8}$ \cite{CPF2}. In 2006, Lin \emph{et al.} presented a scheme for implementing a multiqubit CPF gate by only one step, and showed that $F$ can has its minimum value $F_{min} = Q \approx 0.989$, and the quality factor $Q$ is independent of the variation of coupling rate $g$ \cite{CPF3}. In 2008, Gao \emph{et al.} proposed the scheme to realize CPF gate between two single photons through a single quantum dot in a slow-light photonic crystal waveguide \cite{CPF4}. In 2015, Wei \emph{et al.} proposed the scheme for implementing optical CPF gate by using the single-sided cavity strongly coupled to a single nitrogen-vacancy-center defect in diamond \cite{CPF5}. In the same year, Hao \emph{et al.} presented a scheme to realize the CPF gate between a flying optical photon and an atomic ensemble based on cavity input-output process and Rydberg blockade \cite{CPF6}. In 2016, Hacker \emph{et al.} utilized the strong light–matter coupling provided by a single atom in a high-finesse optical resonator to realize the universal photon–photon CPF gate and achieved an average gate fidelity of ($76.2 \pm 3.6$) percent \cite{CPF7}. In 2020, Kimiaee Asadi \emph{et al.} proposed three schemes to implement the CPF gate mediated by a cavity \cite{CPF8}. In the past years, the CPF gate has already been well studied in the theory, numerical simulation and experiment \cite{CPF1,CPF2,CPF3,CPF4,CPF5,CPF6,CPF7,CPF8,CPF9,CPF10}, which will make our HGSA scheme based on CPF gate more practical and accessible.

In summary, we have presented an efficient scheme for the complete analysis of hyperentangled GHZ state in polarization and time-bin DOFs. In our scheme, the polarization GHZ state is distinguished at the first step. In theory, it is also accessible for us to accomplish the time-bin GHZ state analysis first, and then distinguish the polarization state using the time-bin entanglement. However, in our scheme, the analysis of polarization GHZ state is very simple and efficient through the CPF gate constructed by the cavity-assisted interaction. With the help of the achievable CPF gate and self-assisted mechanism, this scheme can be directly extended to the complete $N$-photon HGSA. Our scheme can save the quantum resource and largely reduce the requirement on nonlinearity, and we hope it will have useful applications in the quantum communication protocols based on polarization and time-bin hyperentanglement.

\end{document}